\documentclass[amsmath,amssymb,floatfix]{revtex4}
\usepackage[pdftex]{graphicx}
\usepackage{amssymb,amsfonts,amsmath}
\usepackage{graphicx,amsmath}
\usepackage{float}

\begin{document}

\title{Can warmer than room temperature electrons levitate above a liquid helium surface ? }

\author{A.D. Chepelianskii$^{(a)}$, Masamitsu Watanabe$^{(b)}$ and Kimitoshi Kono$^{(c,d)}$}
% \footnote{To whom correspondence should be addressed: alexei.chepelianskii@u-psud.fr}}

\affiliation{$(a)$ Laboratoire de Physique des solides, CNRS, Univ. Paris-Sud, Universit\'e Paris-Saclay, LPS-Orsay, 91405 Orsay, France}
\affiliation{$(b)$ RIKEN Nishina Center, Wako, Saitama 351-0198, Japan }
\affiliation{$(c)$ Department of Electrophysics, NCTU, 1001 Ta Hsueh Rd., Hsinchu 300, Taiwan}
\affiliation{$(d)$ RIKEN CEMS, Hirosawa 2-1, Wako, 351-0198 Japan}

% \pacs{73.63.-b,73.50.Pz,74.78.Na}
\begin{abstract}
We address the problem of overheating of electrons trapped on the liquid helium surface by cyclotron resonance excitation. Previous experiments, suggest that electrons can be heated to temperatures up to 1000K more than three order of magnitude higher than the temperature of the helium bath in the sub-Kelvin range. In this work we attempt to discriminate between a redistribution of thermal origin and other out-of equilibrium mechanisms that would not require so high temperatures like resonant photo-galvanic effects, or negative mobilities. We argue that for a heating scenario the direction of the electron flow under cyclotron resonance can be controlled by the shape of the initial electron density profile, with a dependence that can be modeled accurately within the Poisson-Boltzmann theory framework. This provides an  self consistency-check to probe if the redistribution is indeed consistent with a thermal origin. We find that while our experimental results are consistent with the Poisson-Boltzmann theoretical dependence but some deviations suggest that other physical mechanisms can also provide a measurable contribution. Analyzing our results with the heating model we find that the electron temperatures increases with electron density under the same microwave irradiation conditions. This unexpected density dependence calls for a microscopic treatment of the energy relaxation of overheated electrons.
\end{abstract}
\maketitle

Recently electrons on helium have emerged as a promising platform for the study of non-equilibrium physics in ultra high purity two dimensional materials \cite{Kono,Denis0,Denis1,Denis2,PV}. They allowed to explore from a different perspective microwave-induced resistance oscillations and zero-resistance states under microwave irradiation that were discovered in ultra high purity GaAs heterostructures \cite{Zudov1,Mani,Zudov2,Bykov,ZudovGe,ZnO,Theory1,Girvin,IvanPow,TheoryIvan1,TheoryIvan2,Laidet,Zhirov,Dyakonov,Zudovrmp}. Electrons on helium provide a dilute electron system with well-understood disorder potential and weak screening effects thus comparison with theory is simplified compared to GaAs. For example MIRO on electrons on helium strongly depend on the direction of the circular polarisation as expected for non-interacting electrons \cite{heliumcr2017}. In contrast almost no circular polarisation dependence is observed in GaAs \cite{Smet,Kvon,Ganichev}, even if a dependence on the orientation of the linear polarization is present \cite{Mani1,Mani2}. This discrepancy stimulated investigations on the role of edges, contacts and electron-electron interactions \cite{Chepelianskii,Mikhailov,ZRS2018} in GaAs without yet reaching a full understanding. Due to their low density electrons on helium can also display spectacular physical phenomena so far without equivalent in other systems. It has been shown for example, that zero-resistance states on electrons on helium can develop into incompressible phases where the density becomes pinned to a critical value independent of the external confinement \cite{Natcom}; in other cases the electron density can instead become unstable and exhibit time dpendent self-generated oscillations \cite{DenisWatanabe}. 
More generally, the study of non equilibrium phenomena in electrons on helium is strongly connected with the prospect of using Rydberg states \cite{Grimes,Lambert}, created by the interaction of electrons with their image charge inside liquid helium, for quantum computing \cite{Dykman}. So the relaxation times for excited states and the absorption lineshapes have all been carefully investigated \cite{Lea,DenisKono1,DenisKono2,DenisKono3,Lea2,DenisKono4}. 
From this perspective the possibility of creating overheated electrons by excitation of cyclotron resonance that was reported in \cite{DenisCR} is highly interesting since it challenges the view that energy relaxation rates are all relatively fast in the microsecond range \cite{DenisTime} due to two riplon emission processes \cite{Monarkha2} which should prevent an overheating of the electrons. 
In \cite{DenisCR} the temperature under cyclotron resonance driving was estimated from the horizontal spread of the electrons under excitation and from their vertical displacement, both measurements leading a similar temperature of hundreds of Kelvin. However these two redistribution measurements do not allow to characterise the energy distribution of the electrons and to know if it is well described by a hot thermal distribution function. A characterisation of the non equilibrium distribution function is however important as non-thermal phenomena, for example resonant photo-galvanic effects \cite{NSK1,NSK2} or negative mobilities \cite{Kharkov1,Kharkov2} can also lead to a redistribution of electronic charges. In \cite{Closa} the density distribution of overheated electrons on helium was analyzed using Poisson-Boltzmann theory for different equilibrium density profiles predicting a change of direction of the electron flow under CR when the initial density profile of electrons was changed. Here we thus decided to compare these theoretical predictions with experimental results as a way to estimate if the electron energy distribution is indeed close to a thermal one. We find that although thermal excitation is not the only mechanism contributing to electron displacement reasonably good agreement with the theoretical dependence can still be obtained for moderate driving strengths leading to an estimation of electron temperature of several hundred Kelvin consistent with \cite{DenisCR}. We then analyze the dependence of the out-of-equilibrium electron temperature on the electron density finding the surprising conclusion that it increases almost linearly with the electron density. While this work is mainly motivated by the understanding of energy relaxation processes for electrons on helium, we note that the Poisson-Boltzmann equation appear in many other contexts from plasma physics, to electrolytes and biophysics. Indeed one of the authors worked on a very similar problem in the context of hydrophobic electrolytes where the potential trap was created by the backbone of the electrolyte and electrons were replaced by counterions \cite{Electro1,Electro2}. It is thus interesting to perform a quantitative comparison with its predictions in the well defined context of electrons on helium. 
  
\begin{figure}[h]
  \centerline{\includegraphics[clip=true,width=12cm]{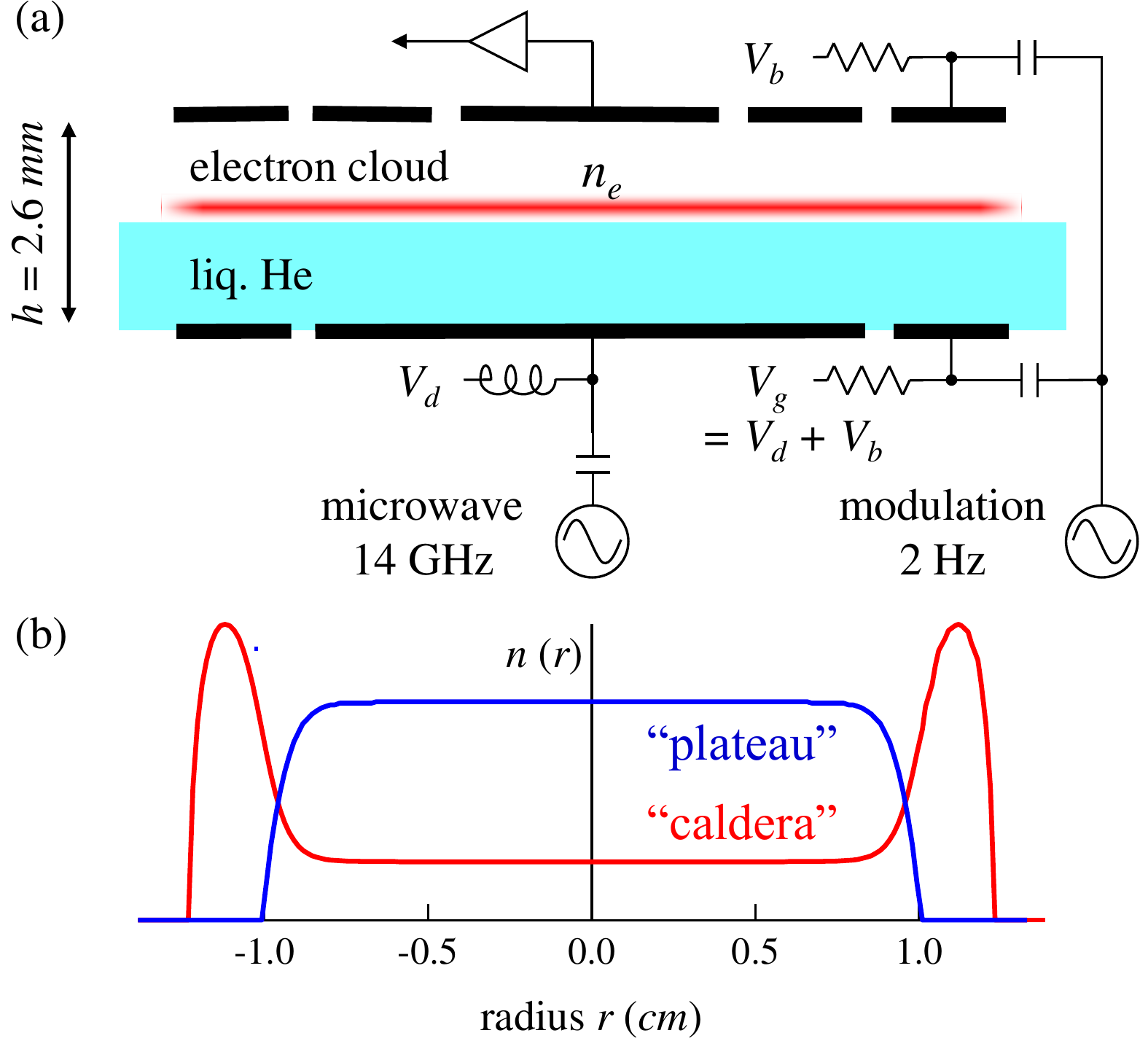}}
\caption{(a) Schematic cross section of the experimental cell with Corbino-disk electrodes. The electrons stick on the liquid helium surface by the attractive force of image charge in the liquid helium and pressing electric field $V_d/h$ created by the voltage difference between top and bottom Corbino electrodes. Electron density profiles are controlled by the bias voltage $V_d$ between the outer guard ring and central electrodes. A bias tee allows to apply a microwave frequency excitation at 14GHz to the bottom disc electrode, and a current amplifier attached to the top disc electrode senses capacitvely changes in the electron density in the center of the cell. For compressibility experiments a low frequency (typically 2Hz) voltage modulation is applied to the guard electrodes through coupling capacitors. Panel (b) shows the typical electron density profiles on the liquid helium surface with Corbino disks that are obtained for positive/negative $V_b$ giving the "caldera"/"plateau" density profiles. 
}
\label{FigCR1}
\end{figure}

The electron density profile inside the sample cell can be controlled by changing the relative voltage between two regions: an outer guard region and the center of the cell (see Fig.~1). Equal densities between the two regions are obtained when the guard voltage $V_g$ is equal to the center voltage $V_d$. When the two voltage are different electrons tend to concentrate in one or the other of the two reservoirs, this leads to "plateau" or "caldera" density profiles which are shown on Fig~1. In the "plateau" profile $V_g < V_d$ and all electrons are attracted towards the center of the cell, while for the "caldera" pattern $V_g > V_d$ the maximum electron density occurs in a ring above the guard electrodes. The change in electronic density between the two regions screens the external electrostatic potential and the electrostatic potential inside the electron cloud is almost constant. This can be seen from the Poisson-Boltzmann mean field theory where the electron density $n$ is connected with the local electrostatic potential $U$ by the relation $n = n_0 \exp(e U/k_B T)$ where $e$ is the absolute value of the electron charge, $k_B T$ is the thermal energy and $n_0$ is a normalization constant for the electron density determined by the total number of electrons trapped in the cloud. 
Thus the potential changes only by amounts of the order of $k_B T$ in cell regions where $n_e$ is non zero or to be accurate not exponentially small. In the caldera pattern the potential barrier to penetrate into the central is thus relatively small and can easily be overcome by heating from cyclotron resonance excitation, as a result electrons will flow towards the center of the cell increasing the density compared to its value when electrons are in thermal equilibrium with the helium bath. On the contrary for the plateau profile, the only potential barrier is on the outer edge of the electron cloud and the main effect of heating is to expand the cloud leading to a drop of the density in the center. The change of the electron density at the center is however smaller compared to the "caldera" profile as the potential barrier at the outer edge is steeper as it is not screened by the electron cloud. The change of the electron density was calculated by numerical simulations of Poisson-Boltzmann equations in \cite{Closa}, it was found that the following approximation for $\Delta n_e$ which is the change of the electron density $n_e$ in the center as function of temperature was valid:
\begin{align}
\Delta n_e = n_e(T) - n_e(T=0) = -\frac{4 \epsilon_0 k_B T [\log n_e(T = 0) - \log n_{av}]}{e^2 h}
\label{eqTemp}
\end{align}
where $\log n_{av}  = \frac{1}{S} \int \log n(T_e = 0) dS$ where the integral is taken of over the electron cloud surface which has an area $S$. As a way to check that Eq.~(\ref{eqTemp}) is consistent with the qualitative picture of the redistribution for "plateau" and "caldera" profiles we can take a simple two state model where electrons are split into central and guard reservoirs of equal surface with respective densities $n_e$ and $n_g$. In this case we find $n_{av} = \sqrt{n_e(0) n_g(0)}$ and Eq.~(\ref{eqTemp}) simplifies to: 
\begin{align}
\Delta n_e \simeq \frac{2 \epsilon_0 k_B T_e}{e^2 h} \log n_g(0)/n_e(0)
\label{eqTemp2}
\end{align}
This approximate expression confirms that we expect $\Delta n_e > 0$ for a "caldera" profile ($n_g(0) > n_e(0)$) and $\Delta n_e < 0$ for a  "plateaus". In the following we report experiments where we check if this sign inversion is present for electrons under cyclotron resonance excitation and then compare our results with quantitative predictions from the Poisson-Boltzmann theory. While no theoretical calculations were made so far to study the dependence of photo galvanic effects on the equilibrium electron cloud density profile, we do not expect a sign reversal in this case as the direction of photo-galvanic currents seems mainly determined by the polarization of the microwave field relative to the helium surface. For comparison between theory and experiments we used Eq.~(\ref{eqTemp}) combined with the zero temperature density profiles obtained from a finite element calculation (see \cite{Natcom}). Note that according to Eq.~(\ref{eqTemp}), $\Delta n_e(V_b)$ is proportional to the electron temperature $T_e$, since all the other parameters appearing in this equation are known the lineshape $\Delta n_e(V_b)/T_e$ is predicted by the heating theory without any fitting parameter. Thus we will sometime normalize data by the best fitting temperature which will allow us to compare our results with the expected theoretical lineshape.

\begin{figure}[h]
  \centerline{\includegraphics[clip=true,width=12cm]{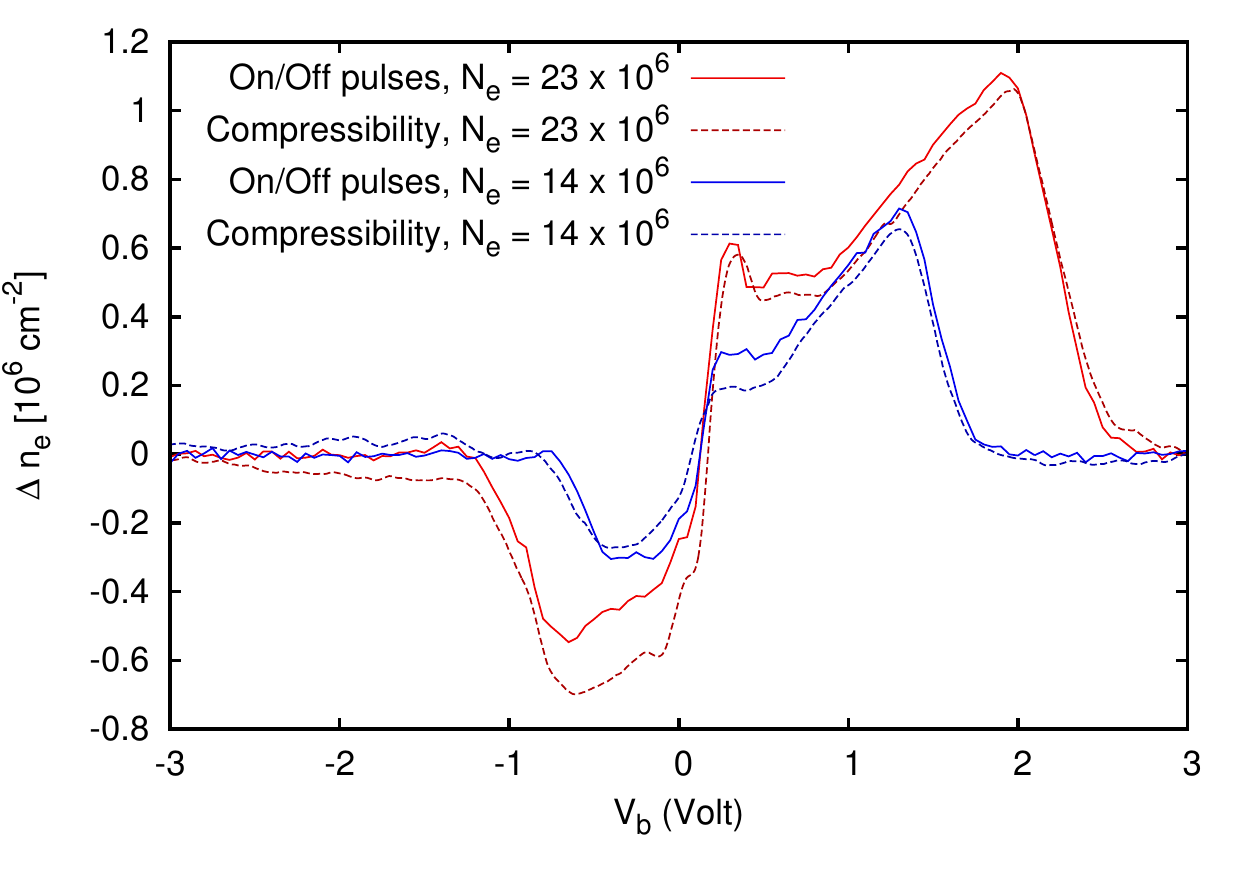}}
\caption{Change in the electron density  $\Delta n_{{\rm e}}$  under cyclotron resonance conditions measured through the photocurrent and compressibility techniques. Cyclotron resonance was excited at a frequency of $\omega = 2 \pi \times 14\;{\rm GHz}$ corresponding to $B = 0.5\;{\rm Tesla}$ at a power of $10$dBm, the holding perpendicular field was $V_{d} = 6\;{\rm Volt}$. The helium bath temperature was fixed to 300mK in all experiments and did not change with cyclotron irradiation. 
}
\label{FigCR2}
\end{figure}

The experiments were performed in a sample cell which has a cylindrical shape and possesses Corbino disk electrodes at the top and bottom of the cell and additional guard electrodes outside of the disk (see Fig.~1), the cell was mounted on a dilution refrigerator with base temperature of around 10mK. The sample cell was inserted in a home-built magnet providing a maximal magnetic field of 1Tesla perpendicular to the helium surface in the sample cell. Liquid helium-4 is inlet into the sample cell and the surface level is adjusted at the center of the top and bottom electrodes. Electrons are deposited onto the helium surface by low power Tungsten filament. 
By applying adequate bias voltage to the bottom-center disk, we can control the shape of the electronic density.
Microwave excitation was applied through the bottom-center Corbino-electrodes which were connected by a cryogenic-coaxial cable to a bias-tee at room temperature. Cyclotron resonance was achieved by applying a microwave excitation at a frequency of 14GHz at a magnetic field of 0.5Tesla.

\begin{figure}[h]
  \centerline{\includegraphics[clip=true,width=12cm]{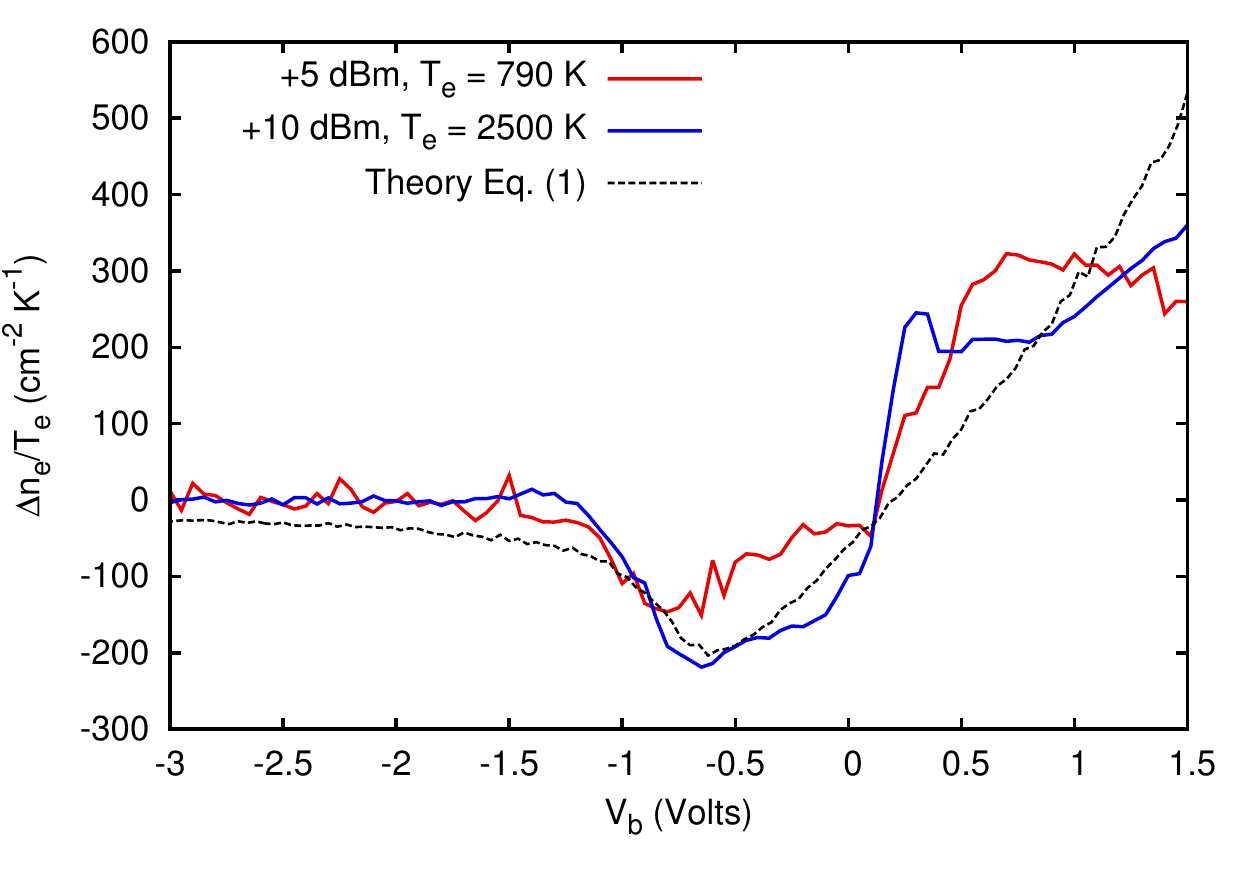}}
\caption{Comparison between Poisson-Boltzmann calculations and the change in the electron density $\Delta n_{{\rm e}}$ under cyclotron resonance measured with the microwave On/Off pulses technique at two microwave powers. The number of trapped electrons was estimated to be $N_e = 23 \times 10^6$ from compressibility experiments without cyclotron irradiation. Experimental values of $\Delta n_e$ are rescaled by the best fitting temperature to ensure the collapse the lineshapes on the predicted theoretical dependence. The obtained temperatures are consistent with an out of equilibrium temperature for cyclotron resonance increasing proportionally to the microwave power.
}
\label{FigCR3}
\end{figure}

Two independent measurement techniques were applied to determine the change of the electron density in the center $n_e$ under CR excitation.  
In the first technique, that can be viewed as a compressibility measurement, we applied a low frequency (2 Hz) modulation voltage $V_{ac}$ on the guard electrode $V_g$. This modulation induces a change of the electron density that was was picked up capacitively from the top-center electrode using a current amplifier using a lock-in amplifier, giving a signal proportional to $d n_e/d V_g$. Integrating this signal over $V_g$ thus allows to reconstruct the dependence $n_e(V_g)$ in equilibrium and under continuous irradiation (CW), subtracting the two results then gives $\Delta n_e$. A detailed description of this approach can be found in \cite{Natcom}. In the other technique, microwaves were applied as On/Off pulses with a frequency of about $1Hz$, the capacitively induced transient currents on the top central Corbino electrode were recorded on an oscilloscope, averaged and integrated to find the induced change in electron density. As opposed to the first technique, this pulsed measurement directly measures $\Delta n_e$. A detailed description of this technique can be found in \cite{PV,DenisCR,Natcom}. We found that for cyclotron resonance the two techniques give the same dependence $\Delta n_e(V_g)$, this contrasts sharply with the redistribution associated to the formation of incompressible states which is hysteretic and thus CW and pulsed excitation can lead to different results \cite{Natcom}. 
Typical experimental results for $\Delta n_e$ are shown on Fig.~2 as a function of $V_b = V_g - V_d$ the voltage difference between the guard and central regions. 
In this experiment $V_d$ was set to $6$Volt and microwave excitation power (at the output of the room temperature generator) was 10dB . In the plateau region for $V_b \le -1.0$Volt, the $\Delta n_e$ sticks to a zero value. This can be interpreted as the guard voltage strongly pushing electrons into the center region, so that the cloud does not expand with cyclotron resonance excitation. Between $V_b = 5.0$ and $V_b = 6.0$, the $\Delta n_e$ has a broad dip with negative value. It is because the electron cloud starts to spread into the guard electrode region and as a result of the weaker repulsive potential from the guard electrodes. At $V_b = 0$ the electron profile crosses over to a caldera profile with a stronger attractive potential towards the guard electrode, we see that as expected for a heating scenario we observe a sign reversal at around this value. Larger $\Delta n_e$ values are then obtained in the caldera region, since the excited electrons can come more easily into the center area from the outer rim of the caldera due to the small potential difference between the two regions. We thus see that qualitative expectations for a heating mechanism for electron redistribution are observed.

\begin{figure}[h]
\begin{tabular}{ll}
a) & b) \\
\includegraphics[clip=true,width=9cm]{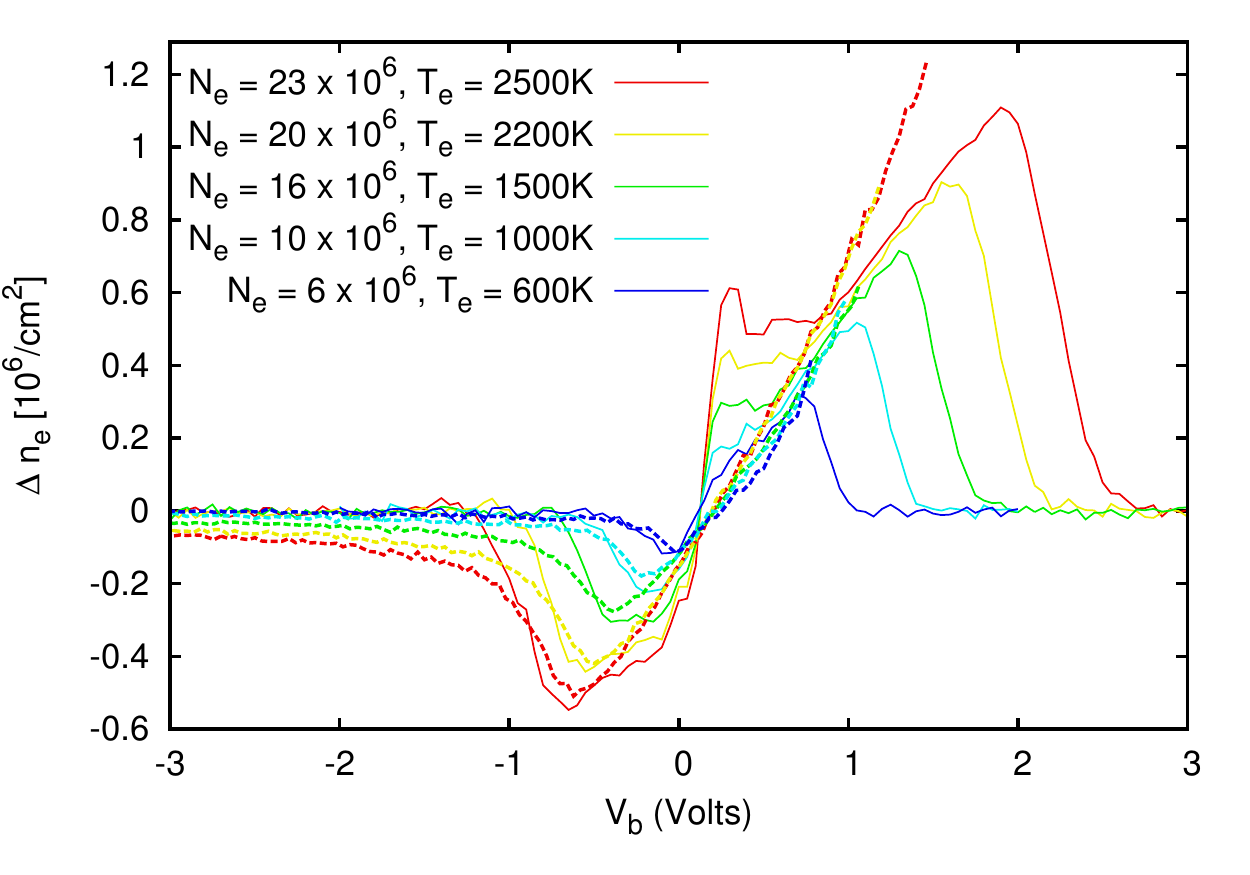} & \includegraphics[clip=true,width=9cm]{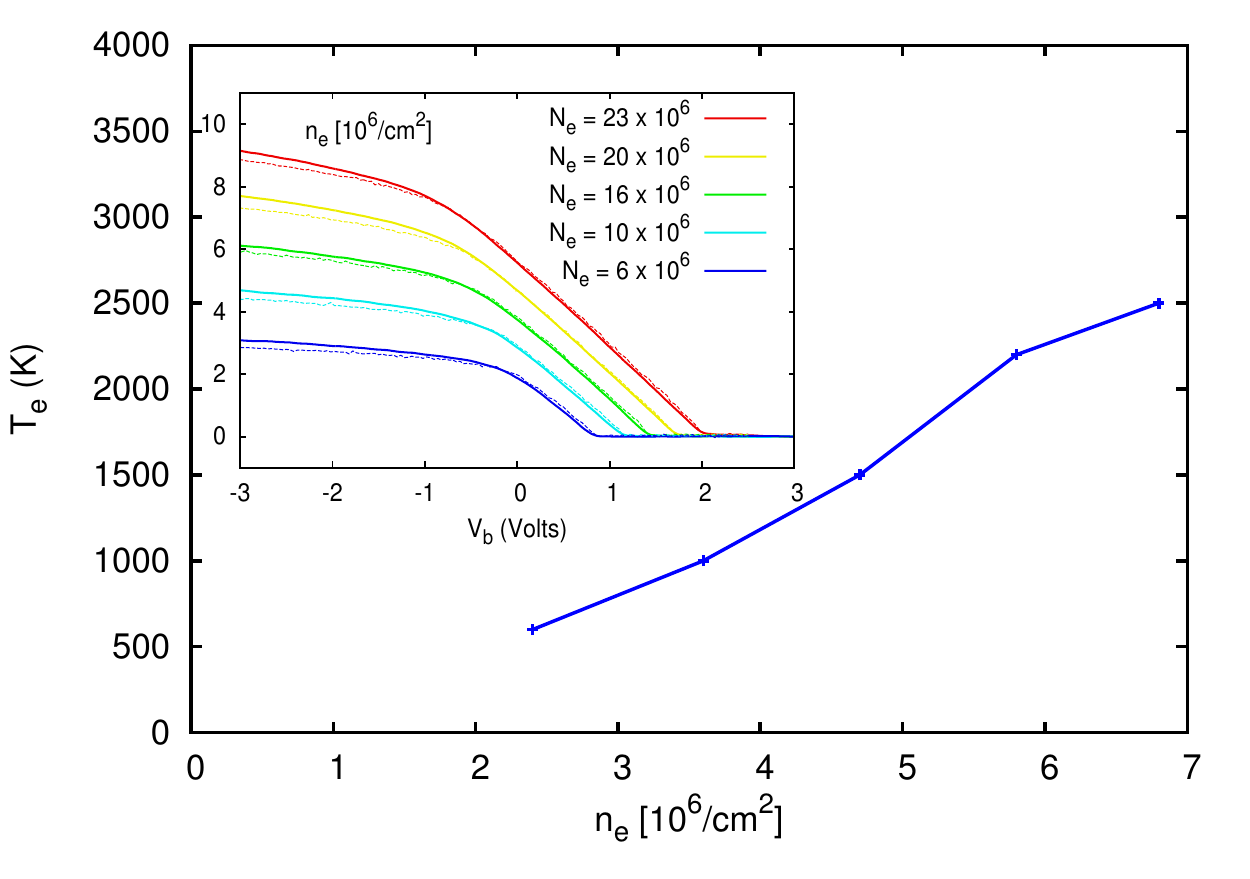}
\end{tabular}
\caption{Panel a) shows the change in the electron density  $\Delta n_{{\rm e}}$ as a function of $V_b$ for different number of electrons trapped in the electron cloud $N_e$ at -10dBm microwave irradiation. The experimental dependence (thick curves) is compared with theoretical predictions from Poisson-Boltzmann theory, the theoretical curves obtained for the best fitting electron temperatures are shown with dashed lines. The obtained temperatures are displayed as a function of the electron density $n_e$ at $V_b = -0.2\;{\rm V}$ in panel b), the temperature increases almost linearly with the electron density.
The inset in panel b) shows $n_e(V_b)$: the dependence of the electron density at the center of cell as function of the bias voltage, it is obtained experimentally from compressibility measurements and compared with calculated finite element curves; both are in good agreement and allow us to find $N_e$ which is needed for the calculations in panel a).
}
\label{FigCR4}
\end{figure}

On Figure 3, we compare the dependence $\Delta n_{{\rm e}}(V_b)$ obtained in the experiments with the theoretical predictions from Eq.~\ref{eqTemp} for two different microwave powers. As explained above we normalized $\Delta n_e{{\rm e}}(V_b)$ by the best fitting temperature (independent of $V_b$) to collapse the different microwave powers on the same theoretical line.
A good agreement was found with a temperature $T = 790\;{\rm K}$ for +5dBm and $T = 2500\;{\rm K}$ for +10dBm, we note that these two values are consistent with a temperature increasing linearly with microwave power. While the agreement is not perfect the overlap between the lineshapes is good for a theory where the only unknown parameter is the vertical scale of the data which is used to determine $T_e$. The region with negative $\Delta n_{{\rm e}}(V_b)$ is reproduced very accurately while stronger deviations appears for positive $\Delta n_{{\rm e}}$. We think that this deviation occurs because the microwave field that excites cyclotron resonance is not homogeneous in the sample cell and is expected to have a maximum at the boundary between the guard and central regions for our excitation geometry. Thus the temperature under cyclotron resonance can vary with $V_b$ in the experiment because the number of electrons excited by cyclotron resonance changes and a perfect overlap between the lineshapes is thus not expected. For this reason we have restricted the range of the comparison to $V_b \le 1.5\;{\rm Volt}$. At higher excitation power a qualitatively different feature appears at $V_b = 0$ which seem to be some sort of resonance between the density of the two reservoirs. The results of Fig.~3 thus support the conclusions of \cite{DenisCR} that electrons on helium can be overheated to temperatures up to $2500\;{\rm K}$ several order of magnitude higher than the temperature of the liquid helium which stayed at 300mK in our experiments. Indeed the sign reversal as function of the bias $V_b$ and the overall lineshape of $\Delta n_{{\rm e}}(V_b)$ are in good agreement with predictions for hot electrons in the framework of Poisson-Boltzmann theory. On the other hand the peak at $V_b = 0$ which appears at the highest power is not expected in the heating theory and may be a manifestation of a redistribution driven by photo-galvanic/negative resistance effects that would not require high electron temperatures.

To characterize the dependence of the electron temperature on the electron density we measured the dependence $\Delta n_{{\rm e}}(V_b)$ at the highest microwave power for different number of electrons $N_e$ trapped in the cloud. The results are presented on Fig.~4.a) with the lineshape obtained with the best fitting temperature from Eq.~(1). We see that the peak at $V_b = 0$Volt develops only when $N_e$ is sufficiently large, and at smaller $N_e$ we recover a dependence which is very close to that expected from Poisson-Boltzmann theory. It is possible that at higher densities (for a fixed external confinement potential) some magnetoplasmon modes couple to cyclotron resonance explaining this result. We note that similar resonances at $V_b = 0$ were also observed for the excitation of Rydberg levels leading to incompressible behavior. To extract the dependence of the electron temperature $T_e$ on the electron density we used the central density for the "plateau" profiles close to the "plateau"-"caldera" transition, since it seems that it is in this region ($-1{\rm Volt} \le V_b \le 0$) that the agreement with the theoretical lineshape is most accurate. Our results for the temperature versus density dependence are shown on Fig.~4.b where for the horizontal density axis we took the values of $n_e$ at $V_b = -0.5\;{\rm Volt}$, the temperatures on the vertical axis were taken from Fig.~4.a. These results leads to an almost linear dependence which is rather surprising. Indeed we expect the temperature under cyclotron-resonance to be approximately independent of the electron density, since both the absorbed microwave power and the power transferred to the helium bath through riplon emission should both be proportional to the electron density which should cancel out from the steady temperature under irradiation. 

\begin{figure}[h]
\begin{tabular}{ll}
a) & b) \\
\includegraphics[clip=true,width=9cm]{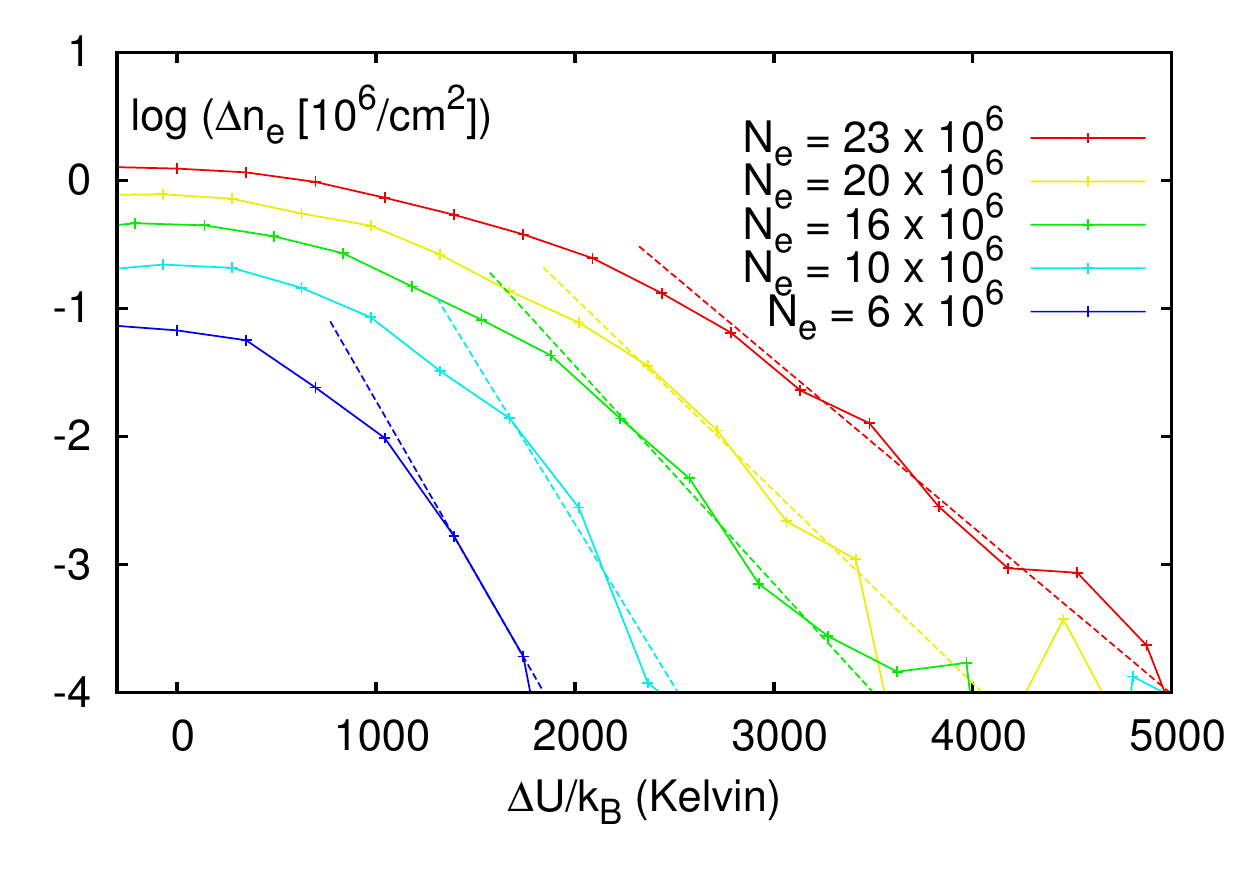} & \includegraphics[clip=true,width=9cm]{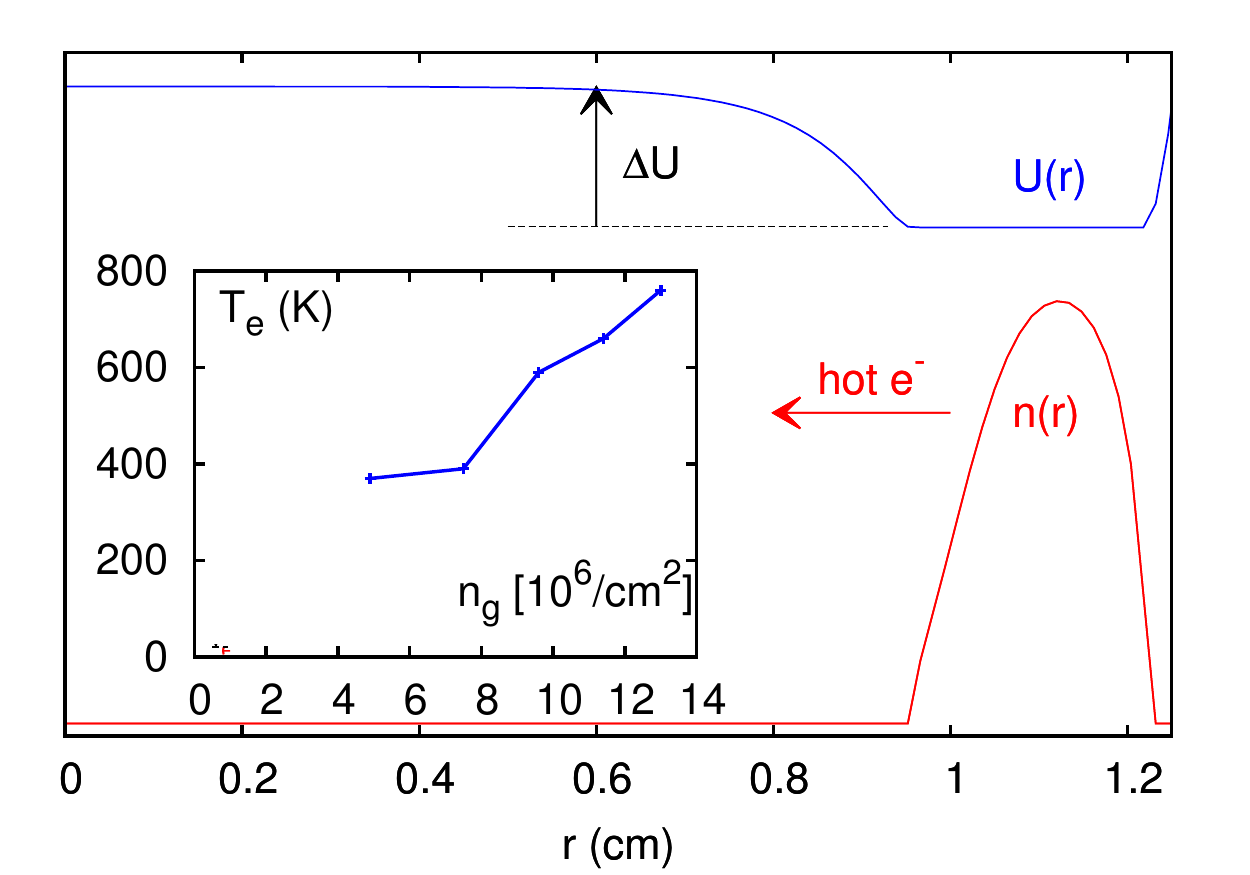}
\end{tabular}
\caption{Panel a) shows $\log \Delta n_{{\rm e}}$  for positive $V_b > V_c$ for which the center of the cell is depleted, the horizontal axis is the energy barrier $\Delta U$ that electrons have to overcome to reach the center of the cell from which the photo-current is detected, the typical density and potential energy profiles in this configuration are shown in panel b). 
The quantitative values for $\Delta U$ were obtained from $V_b$ using finite elements simulations as discussed in the main text. The data is consistent with an exponential decay, the slope of this decay allows to estimate the electron temperature in complementary way compared to Fig.~4. The obtained temperatures are shown on Panel b) as function of the mean electron density $n_g$ above the guard electrodes. Even if the obtained temperatures are lower (presumably because of the less efficient coupling of the microwave irradiation to the electron cloud) they are also consistent with a linear increase as in Fig.~4.
}
\label{FigCR5}
\end{figure}

Aiming to confirm that temperature increases with electron density, we performed a complimentary analysis of the redistribution data from Fig.~4.a. Instead of looking at the charge redistribution at the transition between the "caldera" and "plateau" density profiles, we focused on $\Delta n_e$ values at high $V_b$ where electrons are concentrated in the guard forming a "ring" density profile. Typical density $n(r)$ and potential energy $U(r)$ profiles for the "ring" electron cloud configuration are shown on Fig.~5.b. Due to the attractive potential that confine electrons above the guard electrode, the electrons have to overcome potential energy barrier to reach the center of the cell where pick up currents from the electron redistribution are detected. The height of the potential energy barrier $\Delta U$ can be controlled by changing the bias voltage $V_b$, a finite elements analysis shows that $\Delta U$ depends linearly on $V_b$ following the approximate relation $\Delta U \simeq 0.6 e (V_b - V_c)$ where $V_c$ is the bias voltage value at which the center is depleted $n_e = 0$. For hot electrons we expect that the amount of displaced charge under cyclotron irradiation will follow an activation law $\Delta n_e \propto \exp[-\Delta U/(k_B T_e)]$, from this relation the electron temperature can then be estimated by plotting $\log \Delta n_e$ as function of $\Delta U$. We checked that this procedure is justified within the Poisson-Boltzmann theory by applying it to the $\Delta n_e(V_b)$ dependence obtained from the numerical simulations of \cite{Closa} confirming that it gives the correct electron-temperature. The dependence $\log \Delta n_e$ on the activation energy for the experimental data of Fig.~4.a. is shown on Fig.~5.a., the data is consistent with an activated behavior which can be seen as a strong indication that at least the high energy tail of the electron distribution function is well described by an effective temperature. The extracted temperature under driving is presented in the inset of Fig.~5.b. as function of the mean density in the guard $n_g$, since the electron cloud has the shape of a "ring" with all electrons localized in the guard without cyclotron resonance excitation $n_g$ seems us to be the relevant density parameter in this case. While the temperatures are around a factor 3 lower than in Fig.~4.a, this can be due to the inefficient excitation of cyclotron resonance for the "ring" density profile, and additional experiments with a more homogeneous microwave distribution would be needed to compare the temperatures obtained with the two methods. We note however that the electron temperature still seems to increase linearly with the electron density even for this activation energy analysis.

We now  briefly discuss the possible mechanisms for this unusual dependence of the electron temperature on the electron density. 
Cyclotron resonance narrowing at high densities can lead to an increase of electron temperature because more power is absorbed by a sharper resonance, however at the extremely high temperatures to which electrons seem to be excited the plasma parameter (ratio between Coulomb energy and kinetic energy) will be very small and this narrowing is not expected to occur. Maybe the opposite situation holds instead, at low density overheated electrons could have a very narrow cyclotron resonance since they are only weakly coupled to the helium surface. A very small detuning from cyclotron resonance would then limit the absorbed power. In the limit of larger densities, this narrowing could be suppressed by electron-electron scattering leading to higher temperatures. Careful measurements of the cyclotron resonance linewidth for different excitation powers and electron densities would be needed to probe this scenario. Finally it is possible that many body effects significantly renormalize the cooling power per electron from the riplon bath leading to a reduction of the cooling power at higher densities. Microscopic cooling power calculations for overheated electrons would be needed to understand the origin of the very high temperature that seem to be reached in cyclotron resonance experiments as well as their density dependence.

To summarize, we described some properties of the electron density redistribution that are expected for heating by cyclotron resonance excitation showing that the direction of the electron flow changes sign at the transition from the "plateau" to the "caldera" density profile. The solution of Poisson-Boltzmann equations provides an analytic dependence describing this sign reversal as function of the bias voltage between guard and central reservoirs which controls the transition between the ``plateau'' and ``caldera'' density profiles. We compared this theory with our measurements, obtaining a good agreement which suggests that electrons can indeed become warmer than room temperature even if they are floating above a liquid helium surface. Surprisingly we also found that the temperature of electrons was an increasing function of the density for fixed irradiation conditions. To confirm this result we performed an activation energy analysis measuring the population of electrons that are able to overcome a potential barrier of varying amplitude under irradiation. This leads us to similar conclusions on the density dependence. Our works calls for a more detailed theoretical analysis of energy relaxation processes for highly excited Rydberg states which can make this overheating possible and that could explain this unexpected density dependence.


\begin{thebibliography}{99}

\bibitem{Kono} Y. P. Monarkha and K. Kono
\newblock {\it Two-Dimensional Coulomb Liquids and Solids,}
\newblock {\em Springer-Verlag, Berlin}, (2004)

\bibitem{Denis0} D. Konstantinov and K. Kono
\newblock {\it Novel Radiation-Induced Magnetoresistance Oscillations in a Nondegenerate Two-Dimensional Electron System on Liquid Helium}
\newblock {\em Phys. Rev. Lett.} {\bf 103}, 266808 (2009)

\bibitem{Denis1} D. Konstantinov and K. Kono
\newblock {\it Photon-Induced Vanishing of Magnetoconductance in 2D Electrons on Liquid Helium}
\newblock {\em Phys. Rev. Lett.} {\bf 105}, 226801 (2010)

\bibitem{Denis2} R. Yamashiro, L. V. Abdurakhimov, A. O. Badrutdinov, Yu. P. Monarkha, and D. Konstantinov
\newblock {\it Photoconductivity Response at Cyclotron-Resonance Harmonics in a Nondegenerate Two-Dimensional Electron Gas on Liquid Helium}
\newblock {\em Phys. Rev. Lett.} {\bf 115}, 256802 (2015).

\bibitem{PV}  Denis Konstantinov Alexei Chepelianskii and Kimitoshi Kono
\newblock {\it Photoconductivity Response at Cyclotron-Resonance Harmonics in a Nondegenerate Two-Dimensional Electron Gas on Liquid Helium}
\newblock {\em  J. Phys. Soc. Jpn.} {\bf 81}, 093601 (2012)

\bibitem{Zudov1} Zudov, M.A., Du, R.R., Simmons, J.A. \& Reno, J.L.
\newblock {\it Shubnikov–de Haas-like oscillations in millimeterwave photoconductivity 
in a high-mobility two-dimensional electron gas.}
\newblock {\em Phys. Rev. B} {\bf 64}, 201311(R) (2001).

\bibitem{Mani} Mani, R.G. \& {\it et al.}.
%Ramesh G. Mani*†, Ju ̈rgen H. Smet†, Klaus von Klitzing†,
%Venkatesh Narayanamurti*‡, William B. Johnson§ & Vladimir Umansky
\newblock {\it Zero-resistance states induced by
electromagnetic-wave excitation in GaAs/AlGaAs heterostructures.}
\newblock {\em Nature} {\bf 420}, 646--650 (2002).

\bibitem{Zudov2} Zudov, M.A., Du, R.R., Pfeiffer, L.N. \& West, K.W.
\newblock {\it Evidence for a new dissipationless effect in 2D electronic transport.}
\newblock {\em Phys. Rev. Lett.} {\bf 90}, 045807 (2003).

\bibitem{Bykov} A. A. Bykov, 
\newblock {\it Microwave-induced magnetic field state with zero conductivity in GaAs/AlAs Corbino disks and hall bars}
\newblock JETP Lett. 87, 551 (2008).

\bibitem{ZudovGe} M. A. Zudov, O. A. Mironov, Q. A. Ebner, P. D. Martin, Q. Shi, \& D. R. Leadley
\newblock {\it Observation of microwave-induced resistance oscillations in a high-mobility two-dimensional hole gas in a strained Ge/SiGe quantum well}
\newblock {\em Phys. Rev. B} {\bf 89}, 125401 (2014).

\bibitem{ZnO} D. F. Kärcher, A. V. Shchepetilnikov, Yu. A. Nefyodov, J. Falson, I. A. Dmitriev, Y. Kozuka, D. Maryenko, A. Tsukazaki, S. I. Dorozhkin, I. V. Kukushkin, M. Kawasaki, and J. H. Smet
\newblock {\it Observation of microwave induced resistance and photovoltage oscillations in MgZnO/ZnO heterostructures}
\newblock {\em Phys. Rev. B} {\bf 93}, 041410(R) (2016).


\bibitem{Theory1} V.I. Ryzhii, Sov. Phys. Solid State 11, 2078 (1970).

\bibitem{Girvin} A. C. Durst, S. Sachdev, N. Read, and S. M. Girvin, 
\newblock {\it Radiation-induced magnetoresistance oscillations in a 2D electron gas}
\newblock Phys. Rev. Lett. {\bf 91}, 086803 (2003).

\bibitem{IvanPow} I.A. Dmitriev, A.D. Mirlin and D.G. Polyakov
\newblock {\it Oscillatory ac conductivity and photoconductivity of a two-dimensional electron gas:
Quasiclassical transport beyond the Boltzmann equation}
\newblock Phys. Rev. B {\bf 70}, 165305 (2004)

\bibitem{TheoryIvan1} I. A. Dmitriev, M. G. Vavilov, I. L. Aleiner, A. D. Mirlin, and D. G. Polyakov
\newblock {\it Theory of microwave-induced oscillations in the magnetoconductivity of a two-dimensional electron gas}
\newblock {\em Phys. Rev. B} {\bf 71}, 115316 (2005).

\bibitem{TheoryIvan2} I. A. Dmitriev, A. D. Mirlin, and D. G. Polyakov
\newblock {\it Theory of Fractional Microwave-Induced Resistance Oscillations}
\newblock {\em Phys. Rev. Lett.}  {\bf 99}, 206805 (2007)

\bibitem{Laidet} A. D. Chepelianskii, J. Laidet, I. Farrer, H. E. Beere, D. A. Ritchie, and H. Bouchiat
\newblock {\it Enhancement of edge channel transport by a low-frequency irradiation.}
\newblock {\em Phys. Rev. B} {\bf 86}, 205108 (2012).

\bibitem{Zhirov} Zhirov, O.V., Chepelianksii, A.D.  \& Shepelyansky, D.L.
\newblock {\it Towards a synchronization theory of microwave-induced zero-resistance states.}
\newblock {\em Phys. Rev. B} {\bf 88}, 035410 (2013).

\bibitem{Dyakonov} Beltukov, Y.M.  \& Dyakonov, M.I.
\newblock {\it Microwave-induced resistance oscillations as a classical memory effect.}
\newblock {\em Phys. Rev. Lett.} {\bf 116}, 176801 (2016).

\bibitem{Zudovrmp} Dmitriev, I.A., Mirlin, A.D., Polyakov, D.G.  \& Zudov, M.A.
\newblock {\it Nonequilibrium phenomena in high Landau levels.}
\newblock {\em Rev. Mod. Phys.} {\bf 84}, 1709-1763 (2012).


\bibitem{heliumcr2017} Zadorozhko, A.A., Monarkha Yu.A., Konstantinov D.
\newblock {\it Circular-polarizetion-dependent study of microwave-induced conductivity oscillations 
in a two-dimensional electron gas on liquid helium.}
\newblock {\em Phys. Rev. Lett.}{\bf 120}, 046802 (2017).

\bibitem{Smet} Smet, J.H.  \& {\it et al.}.
% J. H. Smet, B. Gorshunov,  C. Jiang,  L. Pfeiffer,  K. West,  V. Umanksy,  M. Dressel, 
% R. Meisels,  F. Kuchar,  and K. von Klitzing 
\newblock {\it Circular-polarization-dependent study of the microwave photoconductivity
in a two-dimensional rlectron system.}
\newblock {\em Phys. Rev. Lett.} {\bf 95}, 116804 (2005).

\bibitem{Kvon} Herrmann, T.  \& {\it et al.}
%T. Herrmann,  I. A. Dmitriev,  D. A. Kozlov,  M. Schneider,  B. Jentzsch,  Z. D. Kvon,  P. Olbrich, 
%V. V. Bel’kov, A. Bayer,  D. Schuh,  D. Bougeard,  T. Kuczmik,  M. Oltscher,  
%D. Weiss,  and S. D. Ganichev 
\newblock {\it Analog of microwave-induced resistance oscillations induced 
in GaAs heterostructures by terahertz radiation.}
\newblock {\em Phys. Rev. B} {\bf 94}, 081301(R) (2016).

\bibitem{Ganichev} Herrmann, T.  \& {\it et al.}
%T. Herrmann,  Z. D. Kvon, I. A. Dmitriev,  D. A. Kozlov,  B. Jentzsch,  M. Schneider,  L. Schell, 
%V. V. Bel’kov,  A. Bayer,  D. Schuh,  D. Bougeard, T. Kuczmik,  M. Oltscher, 
% D. Weiss,  and S. D. Ganichev
\newblock {\it Magnetoresistance oscillations induced by high-intensity terahertz radiation.}
\newblock {\em Phys. Rev. B} {\bf 96}, 115449 (2017).

\bibitem{Mani1} R. G. Mani, A. N. Ramanayak, and W. Wegscheider 
\newblock {\it Observation of linear-polarization-sensitivity in the microwave-radiation-induced 
magnetoresistance oscillations}, 
\newblock {\em Phys. Rev. B} {\bf 84}, 085308 (2011)

\bibitem{Mani2} A. N. Ramanayaka, R. G. Mani, J. Inarrea, and W. Wegscheider
\newblock {\it Effect of rotation of the polarization of linearly polarized microwaves on the radiation-induced magnetoresistance oscillations}
\newblock {\em Phys. Rev. B} {\bf 85}, 205315 (2012)


\bibitem{Chepelianskii} Chepelianskii, A.D.  \& Shepelyansky, D.L.
\newblock {\it Microwave stabilization of edge transport and zero-resistance states.}
\newblock {\em Phys. Rev. B}, {\bf 80}, 241308(R) (2009).

\bibitem{Mikhailov} S. A. Mikhailov
\newblock {\it Theory of microwave-induced zero-resistance states in two-dimensional electron systems}
\newblock {\em Phys. Rev. B}, {\bf 83}, 155303 (2011)

\bibitem{ZRS2018} Chepelianskii, A.D.  \& Shepelyansky, D.L.
\newblock {\it Microwave stabilization of edge transport and zero-resistance states.}
\newblock {\em Phys. Rev. B}, {\bf 97}, 125415 (2018).

\bibitem{Natcom} Chepelianskii, A.D., Watanabe, M., Nasyedkin, K., Kono, K. \& 
Konstantinov, D.
\newblock {\it An incompressible state of a photo-excited electron gas.}
\newblock {\em Nature Comm.} {\bf 6}, 7210 (2015).

\bibitem{DenisWatanabe} D. Konstantinov, M. Watanabe, K. Kono
\newblock {\it Self-Generated Audio-Frequency Oscillations in 2D Electrons with Nonequilibrium Carrier Distribution on Liquid Helium.}
\newblock  {\em J. Phys. Soc. Jpn.} {\bf 82}, 075002 (2013). 

\bibitem{Grimes} C.C.Grimes, T.R.Brown, M.L.Brown and C.L.Zipfel,
\newblock{\it Spectroscopy of electrons in image-potential-induced surface states outside liquid helium }
\newblock { Phys.Rev. B 13, 140 (1976). }

\bibitem{Lambert} D. K. Lambert and P. L. Richards
\newblock{\it Far-infrared and capacitance measurements of electrons on liquid helium}
\newblock { Phys.Rev. B 23, 3282 (1981). }



\bibitem{Dykman} P. M. Platzman1, and M. I. Dykman
\newblock {\it Quantum Computing with Electrons Floating on Liquid Helium}
\newblock {\em Science} {\bf 284}, 1967 (1999).


\bibitem{Lea} E. Collin, W. Bailey, P. Fozooni, P. G. Frayne, P. Glasson, K. Harrabi, M. J. Lea, and G. Papageorgiou
\newblock {\it Microwave Saturation of the Rydberg States of Electrons on Helium  }
\newblock {\em Phys. Rev. Lett} {\bf 89}, 245301 (2002).


\bibitem{DenisKono1} Denis Konstantinov, Hanako Isshiki, Yuriy Monarkha, Hikota Akimoto, Keiya Shirahama, and Kimitoshi Kono
\newblock {\it Microwave-Resonance-Induced Resistivity: Evidence of Ultrahot Surface-State Electrons on Liquid }
\newblock {\em Phys. Rev. Lett} {\bf 98}, 235302 (2007).

\bibitem{DenisKono2} Y. Monarkha, D. Konstantinov and K. Kono
\newblock {\it Microwave absorption saturation and decay heating of surface electrons on liquid helium}
\newblock {\em Low Temperature Physics } {\bf 33}, 718 (2007).

\bibitem{DenisKono3} Denis Konstantinov, M. I. Dykman, M. J. Lea, Yuriy Monarkha, and Kimitoshi Kono
\newblock {\it Resonant Correlation-Induced Optical Bistability in an Electron System on Liquid Helium}
\newblock {\em Phys. Rev. Lett} {\bf 103}, 096801 (2009).

\bibitem{Lea2} E. Collin, W. Bailey, P. Fozooni, P. G. Frayne, P. Glasson, K. Harrabi, and M. J. Lea
\newblock {\it Temperature-dependent energy levels of electrons on liquid helium }
\newblock {\em Phys. Rev. B} {\bf 96}, 235427 (2017).

\bibitem{DenisKono4} M. I. Dykman, K. Kono, D. Konstantinov, and M. J. Lea
\newblock {\it Ripplonic Lamb Shift for Electrons on Liquid Helium}
\newblock {\em Phys. Rev. Lett} {\bf 119}, 256802 (2017)

\bibitem{DenisCR} A. O. Badrutdinov, L. V. Abdurakhimov, and D. Konstantinov
\newblock {\it Cyclotron resonant photoresponse of a multisubband two-dimensional electron system on liquid helium}
\newblock {\em Phys. Rev. B} {\bf 90}, 075305 (2014)

\bibitem{DenisTime} A. O. Badrutdinov, D. Konstantinov, M. Watanabe and K. Kono
\newblock {\it Experimental study of energy relaxation of hot electrons on liquid helium-4}
\newblock {\em EPL} {\bf 104}, 47007 (2013)

\bibitem{Monarkha2} Y. P. Monarkha
\newblock {\it Influence of shortwave surface excitations of liquid helium on damping effects in a two-dimensional electron gas}
\newblock {\em Sov. J. Low Temp. Phys.} {\bf 4}, 515 (1978)

\bibitem{NSK1} M. V. Entin and L. I. Magarill
\newblock {\it Surface photocurrent in an electron gas over liquid He subjected to a quantizing magnetic field}
\newblock {\em JETP Letters} {\bf 98}, 744 (2014)

\bibitem{NSK2} M. V. Entin and L. I. Magarill
\newblock {\it Photogalvanic current in electron gas over a liquid helium surface}
\newblock {\em JETP Letters} {\bf 98}, 816 (2014)

\bibitem{Kharkov1}  Y. P. Monarkha
\newblock {\it  Cyclotron-resonance-induced negative dc conductivity in a two-dimensional electron system on liquid helium }
\newblock {\em  Phys. Rev. B } {\bf 91}, 121402 (2015)

\bibitem{Kharkov2}  Y. P. Monarkha
\newblock {\it   Density domains of a photo-excited electron gas on liquid helium}
\newblock {\em  Low Temp. Phys. } {\bf 42}, 441 (2016)

\bibitem{Closa} Fabien Closa, Elie Raphäel and Alexei D. Chepelianskii
\newblock {\it Transport properties of overheated electrons trapped on a helium surface}
\newblock {\em Eur. Phys. J. B} {\bf 87}, 190 (2014)

\bibitem{Electro1} A. Chepelianskii, F. Mohammad-Rafiee, E. Trizac and E. Raphaël 
\newblock {\it On the Effective Charge of Hydrophobic Polyelectrolytes}
\newblock {\em J. Phys. Chem. B} {\bf 113}, 3743 (2009)

\bibitem{Electro2}  A. D. Chepelianskii, F. Closa, E. Raphaël and E. Trizac 
\newblock {\it Strong screening in the plum pudding model}
\newblock {\em } {\bf 94}, 68010 (2011)


\end{thebibliography}
\end{document}